\documentclass[universe,article,submit,moreauthors,pdftex]{Definitions/mdpi} 

\firstpage{1} 
\pubvolume{1}
\issuenum{1}
\articlenumber{0}
\pubyear{2021}
\copyrightyear{2020}
\datereceived{} 
\dateaccepted{} 
\datepublished{} 
\hreflink{https://doi.org/} 

\usepackage{natbib}	

\Title{
Does the GRB duration depend on redshift?
}

\TitleCitation{Does the GRB duration depend on redshift?}

\Author{I. Horvath$^{1,*}$\orcidB{},
I.I. Rácz$^{1}$\orcidE{},
Z. Bagoly$^{2}$\orcidC{},
L.G. Balázs$^{3,4}$\orcidF{} and 
S. Pinter$^{1}$\orcidA{}}

\AuthorNames{Istvan Horvath et. al.}

\AuthorCitation{Horvath, I.; Racz, I.I.; Bagoly, Z.; Balazs L.G.; Pinter, S.}
\address{%
$^{1}$ \quad University of Public Service, Budapest, Hungary\\
$^{2}$ \quad Department of Physics of Complex Systems, E\"otv\"os Lor\'and University, Budapest, Hungary\\
$^{3}$ \quad Department of Astronomy, E\"otv\"os Lor\'and University, Budapest, Hungary\\
$^{4}$ \quad Konkoly Observatory, Research Centre for Astronomy and Earth Sciences, Budapest, Hungary
}

\corres{Correspondence: horvath.istvan@uni-nke.hu}

\abstract {Several hundreds Gamma-ray Burst (GRB) redshift had been
determined so far. One of the other important properties
--besides the distance-- of the GRBs is the duration of the burst.
This article statistically analyses these two important quantities of the phenomena. 
In this paper we map of the two-dimensional distribution and explore some suspicious  areas. 
As it is well known the short GRBs are closer than the others hence we  search for  parts in the Universe where the GRBs duration different from the others.
We also analyse whether there are any range in the duration where the redshifts are differing.
We find some suspicious  areas, however, no other significant region was found than the short GRBs region.}


\begin{document}

\section{Introduction}

Gamma-ray bursts (GRBs) are the most energetic phenomena in the
universe, which can be detected up to very high redshifts ($z = 8.2$ spectroscopically \cite{2009NaturS,2009NaturT} and $z=9$ by photometry \cite{Cucchiara:2011}). 
Although there are indications that there are more GRB groups \cite{hor98,hak00,hor02,hor04,rmw09,kb12,tsu13,zito15,tarno15AA,tarno16NewA,hth18,2019ApJTarn,2022AATarn} other than the classical short and long dichotomy \citep{maz81,nor84}, in this paper we deal only with short (SGRB) and long (LGRB) gamma-ray bursts.
The cosmological origin of gamma-ray bursts (GRBs) is well established (e.g. \citet{MG12}).
Assuming that the Universe exhibits large-scale homogeneity and isotropy, the same is also expected for GRBs. 
GRBs so far the only objects which are sampling the observable Universe as a whole, thus the large-scale angular isotropy of the sky distribution of GRBs has been well studied over the last decades. 
Most of these studies have demonstrated that the sky distribution of GRBs is isotropic \citep{Briggs96,Teg96,bal98,bal99,mesz00,mgc03,vbh08}.

However, there are indications for some large-scale anisotropy in space distribution of quasars \cite{clo12} and GRBs \cite{hhb13,hhb14,hbht15,BalazsRing2015,HSZ20}.
Some GRB subsamples appear to deviate significantly from isotropy. 
\citet{bal98} and \citet{vbh08} reported that the angular distributions of short and long GRBs are different, while  \citet{2017ifs..confE..80R,R_pa_2017,2019MNRAS.486.3027R} analyzed the isotropy of Fermi GRBs according to their properties (duration, fluences, peak fluxes). 

\citet{Cline99} found that the angular distributions of very short GRBs are anisotropic, and 
\citet{mgc03} reported that the short GRB class in general deviates from angular isotropy. 
\citet{mesz00} and \citet{li01} wrote that the angular distribution of intermediate duration ($T_{90}$) GRBs is not isotropic. 
LGRBs are believed to produce from core-collapsed supernovae (SNe) \citep{woo93,paczy98,wb06}
which is supported by observations that some GRBs
associated with SNe,  (e.g., \citet{hjor03,stan03}
). 
This model implies that the LGRB event rate should trace the cosmic star formation rate \citep{tot97,Zhang:2004,mesz06,zh07ChJA}.
They generally occur in faint, blue, low-mass star-forming galaxies \citep{LeFloch:2003,Palmerio:2019}
and also in the bright regions of their hosts 
\citep{Fruc06Natur,blan16,lyman17}

\begin{figure}
	 \centering
     \includegraphics[width=0.7\columnwidth]{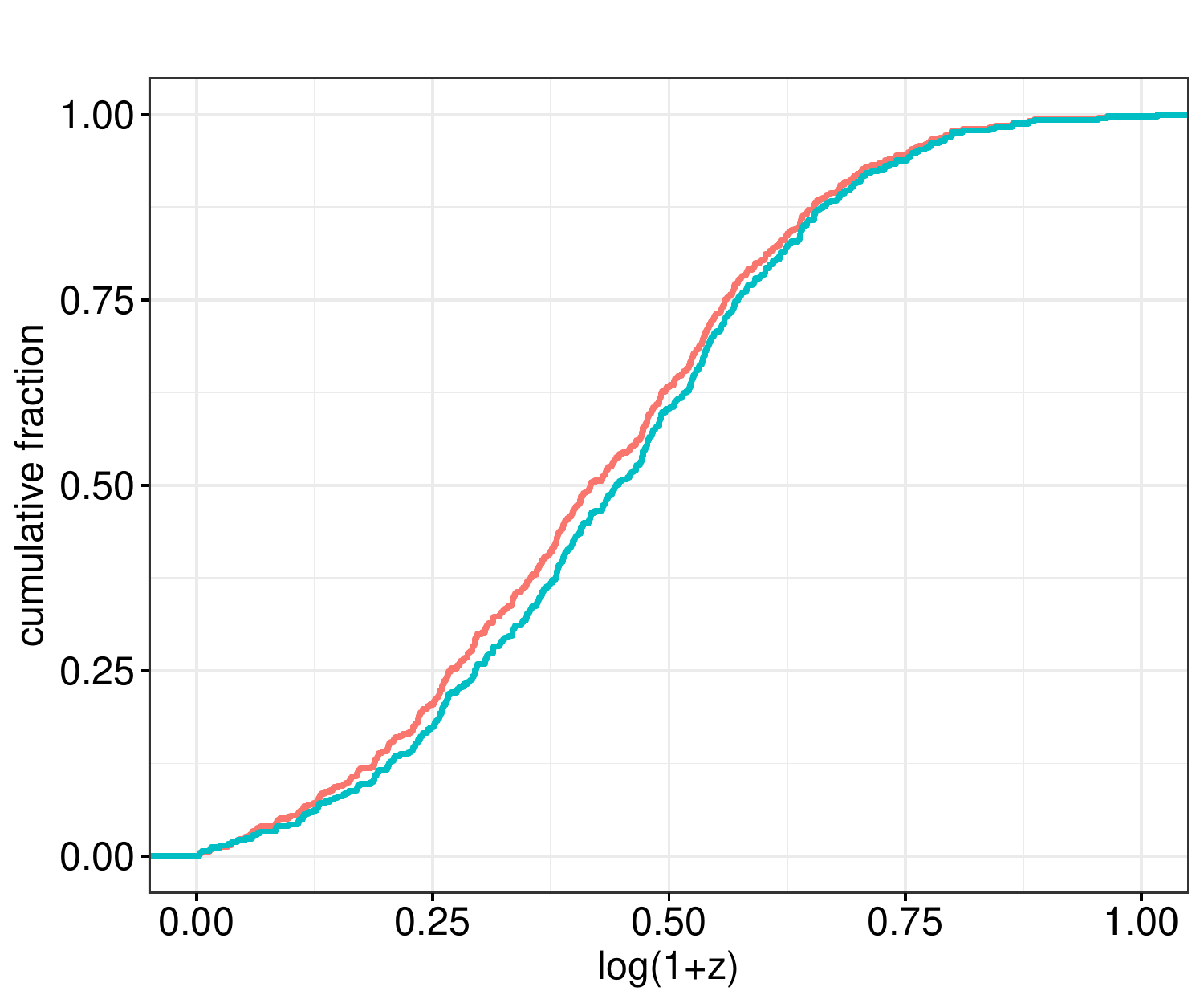}
     \caption{ The redshift cumulative distributions of the 474 bursts (red) and the 421 non-short ($T_{90} \geq 5s$)  GRBs' (blue) which had known redshift and duration.
     }
     \label{fig:fig425comp474}
     \end{figure}

\begin{figure}
 \centering
 \includegraphics[width=0.7\columnwidth]{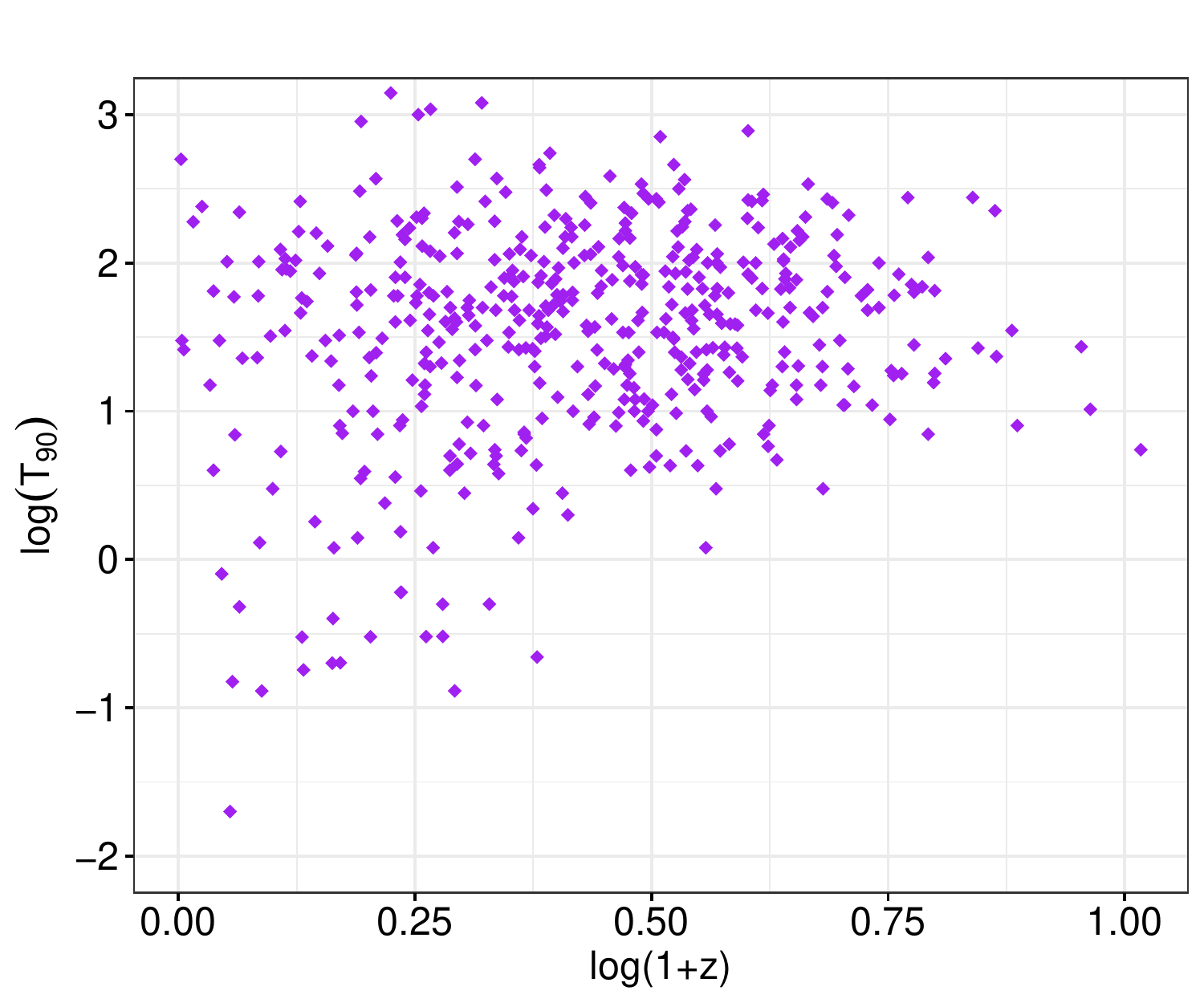}
   \caption{Duration ($T_{90}$) vs. log $(1+z)$ distribution of the 474 GRBs with known redshift and duration.}
  \label{fig:figTvsz}
\end{figure}

\section{ Duration versus redshift distribution analyses }\label{sec:analyses}

Nowadays, nearly five hundred redshifts were observed for GRBs. 
The Caltech GRBOX web-page contains the biggest part of them\footnote{\url{http://www.astro.caltech.edu/grbox/grbox.php} 
}, 
therefore in this analysis we use their data set.
Among these 487 GRBs 474 had duration information as well.
We study these bursts spatial distributions.

On Fig.~\ref{fig:fig425comp474} the cumulative distribution of the redshifts of the 474 GRBs can be seen: one can observe that the $\sim 60\%$ of the GRBs are in the  $\log(1+z)= ( 0.25 - 0.6 )$   range.

\subsection{Comparison with the whole sample redshifts}\label{sec:Tzanal}

\begin{figure}
 \centering
 \includegraphics[width=0.7\columnwidth]{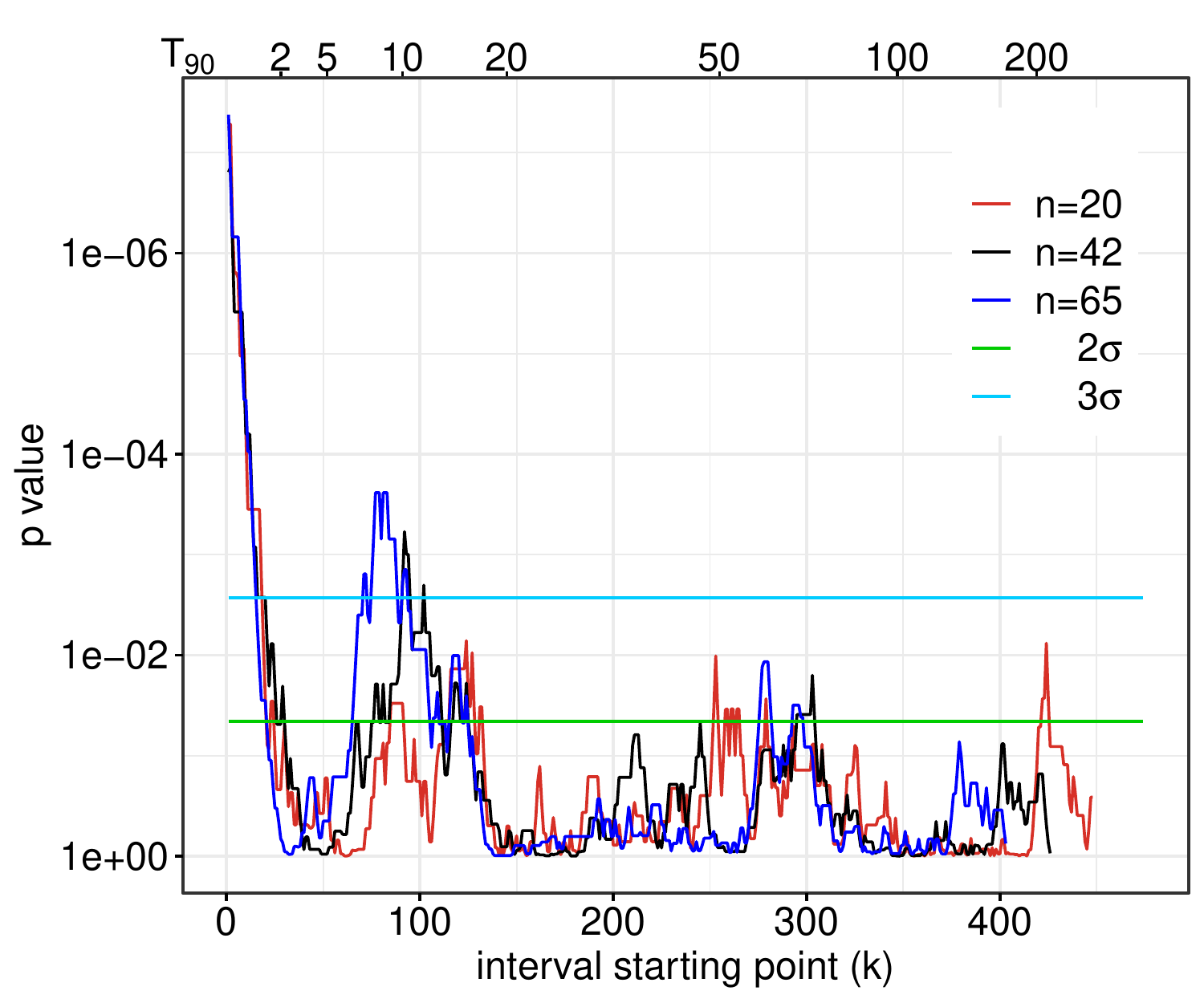}
   \caption{One can order the 474 GRBs according to their $T_{90}$. $n$=20  (red), $n=42$ (black) and $n=65$ (blue) consecutive GRBs were chosen and was compared their $z$ distribution with the complementary $474-n$ GRBs' redshift  distributions.  
   This figure shows the KS test $p$ value as a function of the starting number of the $n$ GRBs. 
   Green (light blue) line marks the 2$\sigma$ (3$\sigma$) significance level.  }
  \label{fig:fig474all}
\end{figure}

Fig.~\ref{fig:figTvsz} shows the redshift vs. duration ($T_{90}$) distribution of the 474 GRBs.
To study whether the redshift distribution depend on duration ($T_{90}$) parameter, one can use several statistical tests.
Here we ordered the GRBs by duration and chose $n$ consecutive ones. This group's redshift distribution was compared with the complementary $474-n$ GRBs' redshift distribution using  Kolmogorov-Smirnov test (KS). (Note: in this paper we use 10 base logarithm, therefore $log$  always means $log_{10}$.)

We compared the redshift distributions starting the group at the $k$th position. We did this process for different group sizes from $n=8$ to $n=99$. As an example Fig.~\ref{fig:fig474all} 
shows the KS $p$ value's dependence of $k$ for $n = 20$, $42$ and $65$, respectively. The green (0.0455) and blue (0.0027) lines show 
2 and 3 sigma significance, respectively.

\subsection{Comparison with the 421 non-short GRBs' redshifts}\label{sec:scre}

From the KS tests results on Fig.~\ref{fig:fig474all} one can see that the biggest deviations are for the shortest bursts. 
It has been well known that these short GRBs are typically closer in distance  
than long GRBs \cite{2006A&A...453..797B,2009AIPC.1133..455J,peram10}. Some believe that 2 seconds is the border between short and long bursts \citep{Briggs96}, however many studies show the border is not that obvious \citep{bal98,bal99,mesz00,zha09}. 

\begin{figure}
	 \centering
	 \includegraphics[width=0.7\columnwidth]{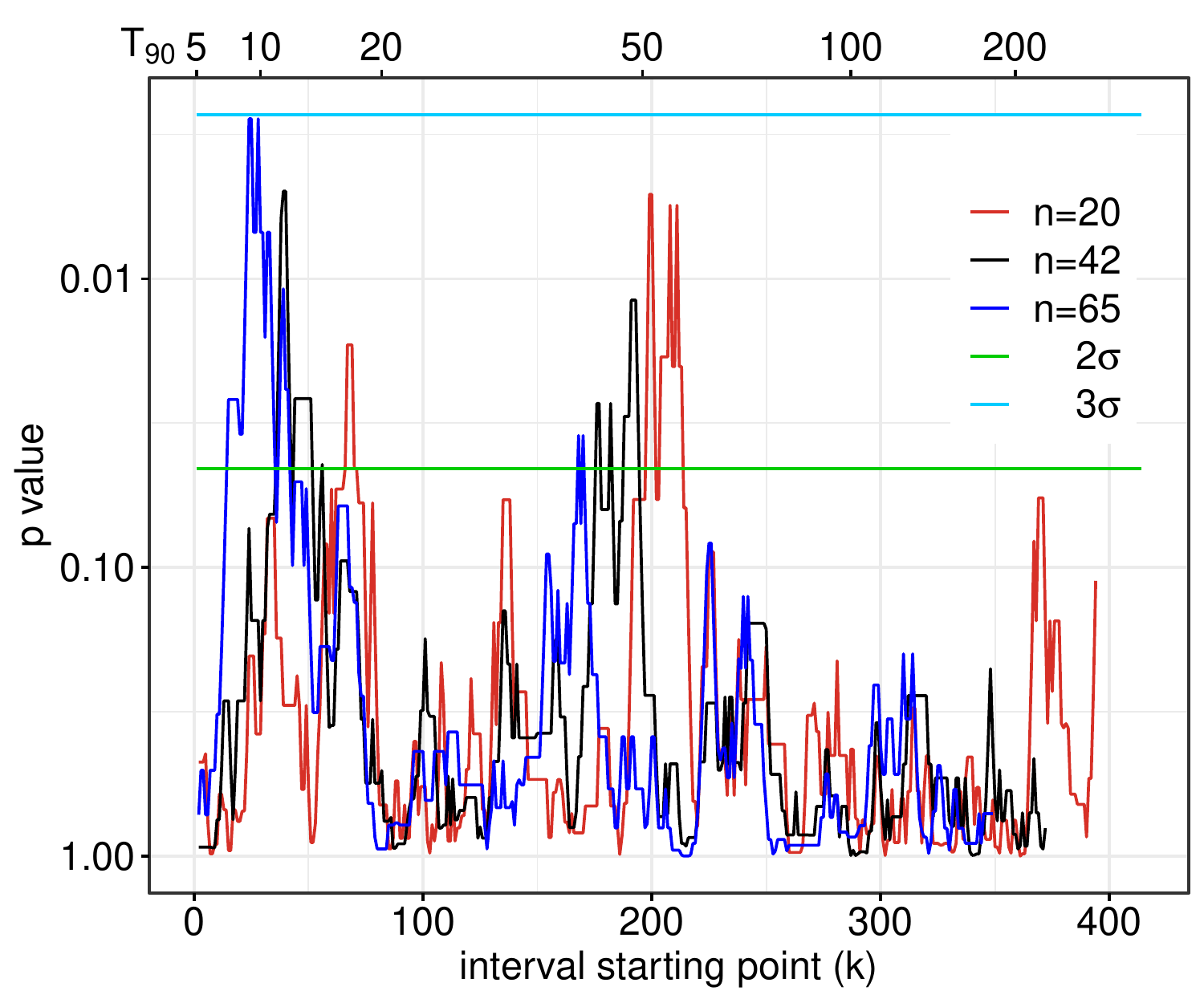}
     \caption{We ordered the 421 non-short GRBs ($T_{90}<5s$) according to their $T_{90}$. $n$=20  (red), $n=42$ (black) and $n=65$ (blue) consecutive GRBs were chosen and was compared their z distribution with the complementary $421-n$ GRBs' redshift  distributions’. This figure shows the KS test $p$  value as a function of a starting number of the $n$ GRBs. 
   Green (light blue) line marks the 2$\sigma$ (3$\sigma$) significance level.}
     \label{fig:fig42520}
\end{figure}

As the different groups' $T_{90}$ distributions are quite wide, we chose the $T_{90}=5.0s$ for this division to make sure there are no short GRBs in the sample.
This means the first 53 GRBs ($T_{90} \leq 5s$) were not used for calculating the distribution of the long GRBs' redshift distribution. 
(For the difference of the 421 long GRBs' distribution and the whole 474 GRBs sample distribution see Fig.~\ref{fig:fig425comp474}.) 
Then we repeated the method which was described in Sect.~\ref{sec:Tzanal}.
We also did this process from $n=8$ to $n=99$. Fig.~\ref{fig:fig42520}  shows the KS $p$ value's dependence of $k$ for $n = 20$, $42$ and $65$ respectively.  The green (0.0455) and blue (0.0027) lines show 2 and 3 sigma significance, resp.

\begin{figure}
	 \centering
	 \includegraphics[width=0.7\columnwidth]{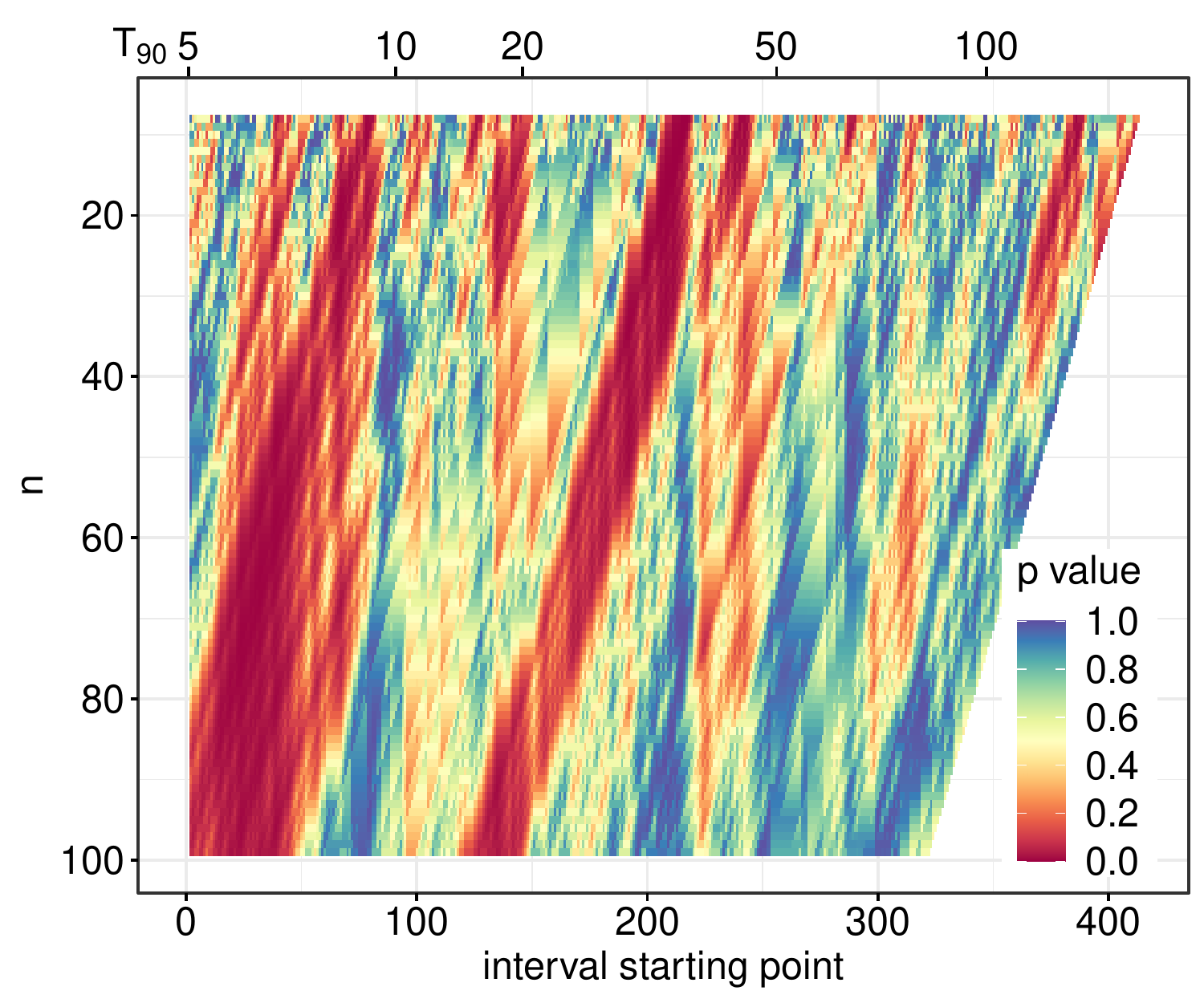}
   \caption{The two parameter (n, duration) KS test $p$ value  contour plot. Using the 421 non-short GRBs. }
  \label{fig:figz0D3sik}
\end{figure}

Fig.~\ref{fig:figz0D3sik} shows the two parameter ($n$, duration) KS $p$ value. Note that the short part was cut from the figures, since $p$ is extremely low in the short duration area (see Fig.~\ref{fig:fig474all}), which also means high significance. This means the short GRBs redshift distribution differ in a very high significance level from the long duration GRBs redshift distribution. We call this part of the GRBs' distribution AREA1.

\begin{figure}
	 \centering
	 \includegraphics[width=0.7\columnwidth]{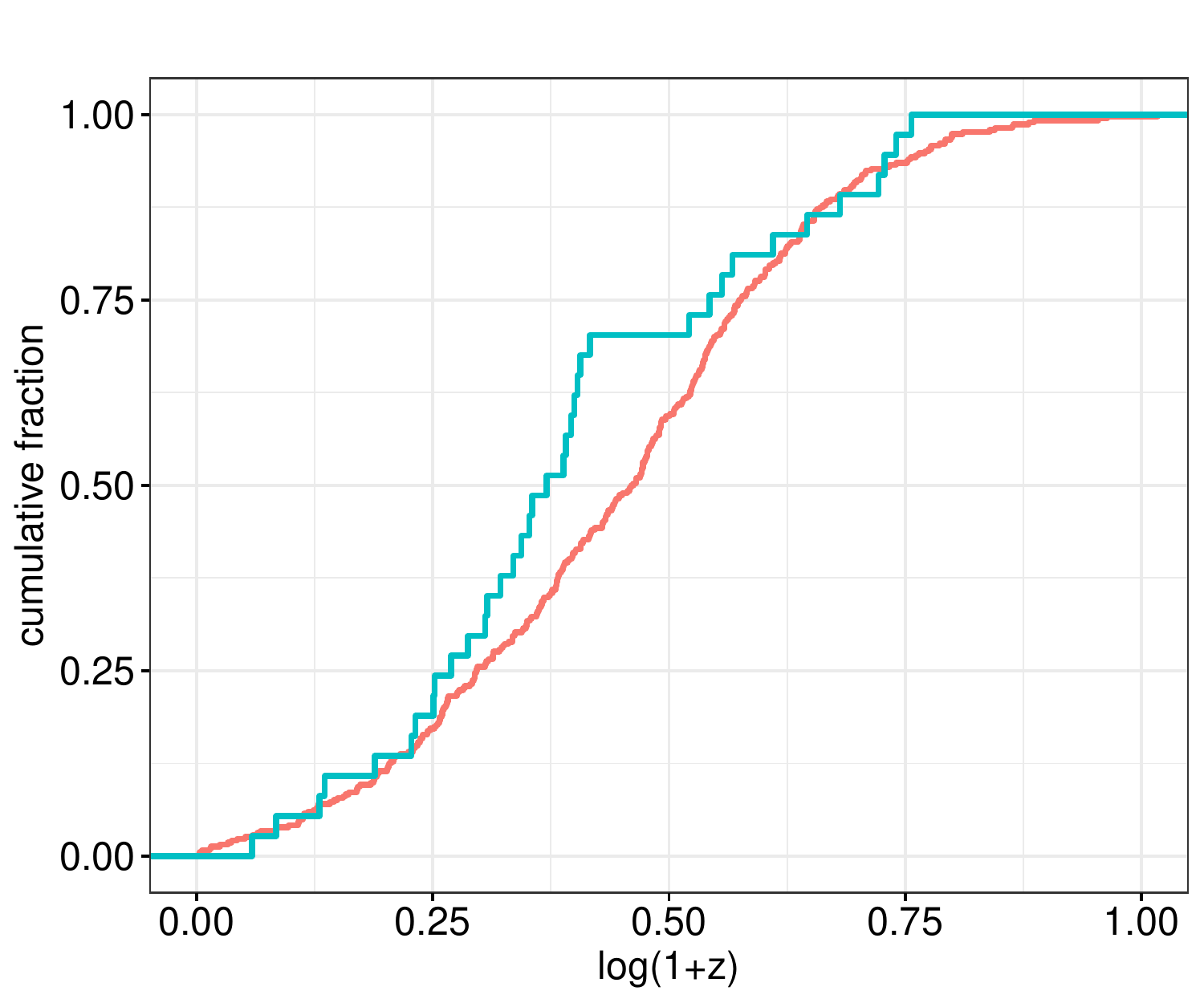}
   \caption{Cumulative redshift distribution of AREA3 GRBs ($49s < T_{90} < 61s$) is marked with blue. Red line is the 421-$n$ GRBs' redshift distribution. These bursts tend to be closer than the others. }
  \label{fig:figarea3}
\end{figure}

\begin{figure}
	 \centering
	 \includegraphics[width=0.7\columnwidth]{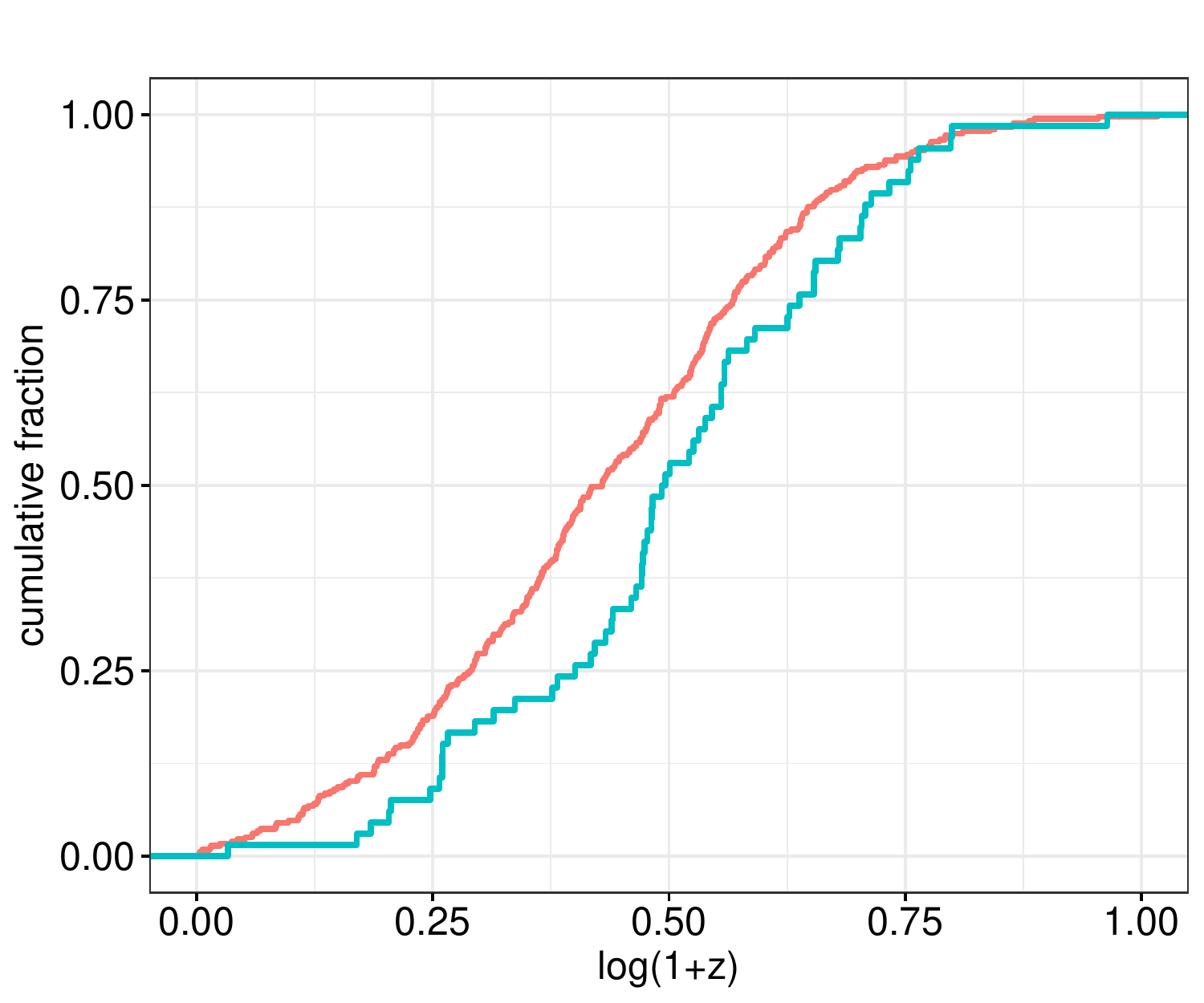}
   \caption{Cumulative redshift distribution of AREA4 GRBs ($9s < T_{90} < 21s$) is marked with blue. Red line is the 421-$n$ GRBs' redshift distribution.  These bursts tend to be farther than the others. }
  \label{fig:figarea4}
\end{figure}

The $p$ value reaches 0.0027 in three area. The aforementioned short duration part (AREA1), the $T_{90}$ (16s, 20s) interval (AREA2) with about $n$ between 12 and 21, and the $T_{90}$ (49s, 61s) interval (AREA3) with about $n$ between 23 and 36. The $p$ value reaches 0.0455 in only four area. The three aforementioned areas and the $T_{90}$ (9s, 21s) interval (AREA4) with about $n$ between 59 and 68.
Fig.~\ref{fig:figarea3}
and Fig.~\ref{fig:figarea4} show the cumulative redshift distribution of AREA3-4.


\subsection{The redshift vs. T\textsubscript{90}  method}\label{sec:point}

One can make a similar analysis by swapping the variables: order the GRBs by redshift, select a redshift interval, then compare the duration distribution of this subsample with the duration distribution of the complementary sample. Since there are few short bursts with redshift bigger than one we omitted the 53 GRBs which had $T_{90} \leq 5$s.
Therefore, we analysed the remaining 421 GRBs. 
Here we ordered the GRBs by redshift and chose the closest, consecutive $n$ GRBs and compared the $n$ closest GRBs' duration distribution with the $421-n$ GRBs' duration distribution, performing the Kolmogorov-Smirnov test (KS).
We repeat this process starting from the $k$th GRB and repeated the process with a block size of $n$ running from 8 to 99.

\begin{figure}
	 \centering
	 \includegraphics[width=0.7\columnwidth]{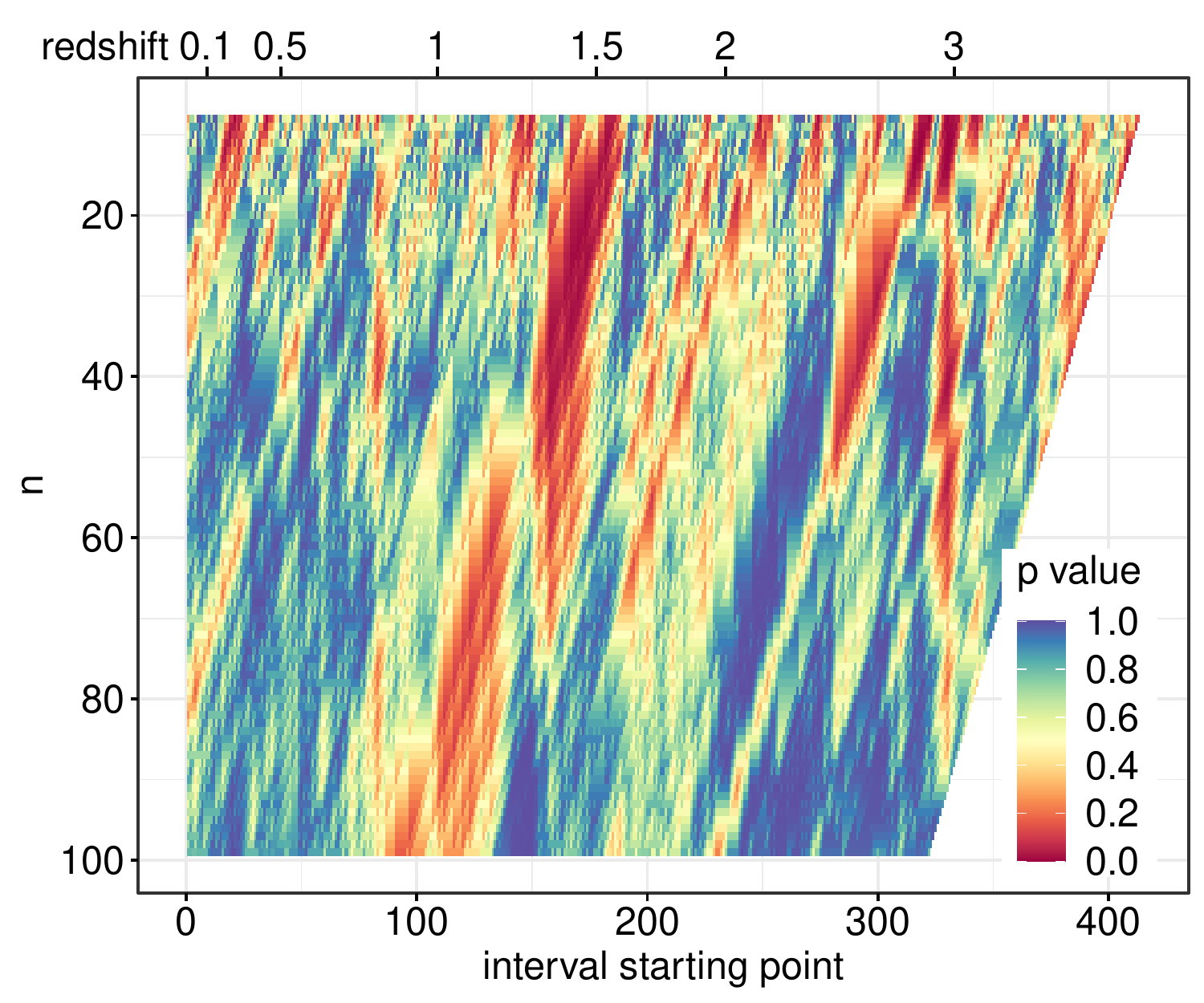}
   \caption{The two parameter (n, redshift) KS $p$ value surface plot. Using the 421 non-short GRBs.}
  \label{fig:figz0D4sik}
\end{figure}

\begin{figure}
 \centering
 \includegraphics[width=0.7\columnwidth]{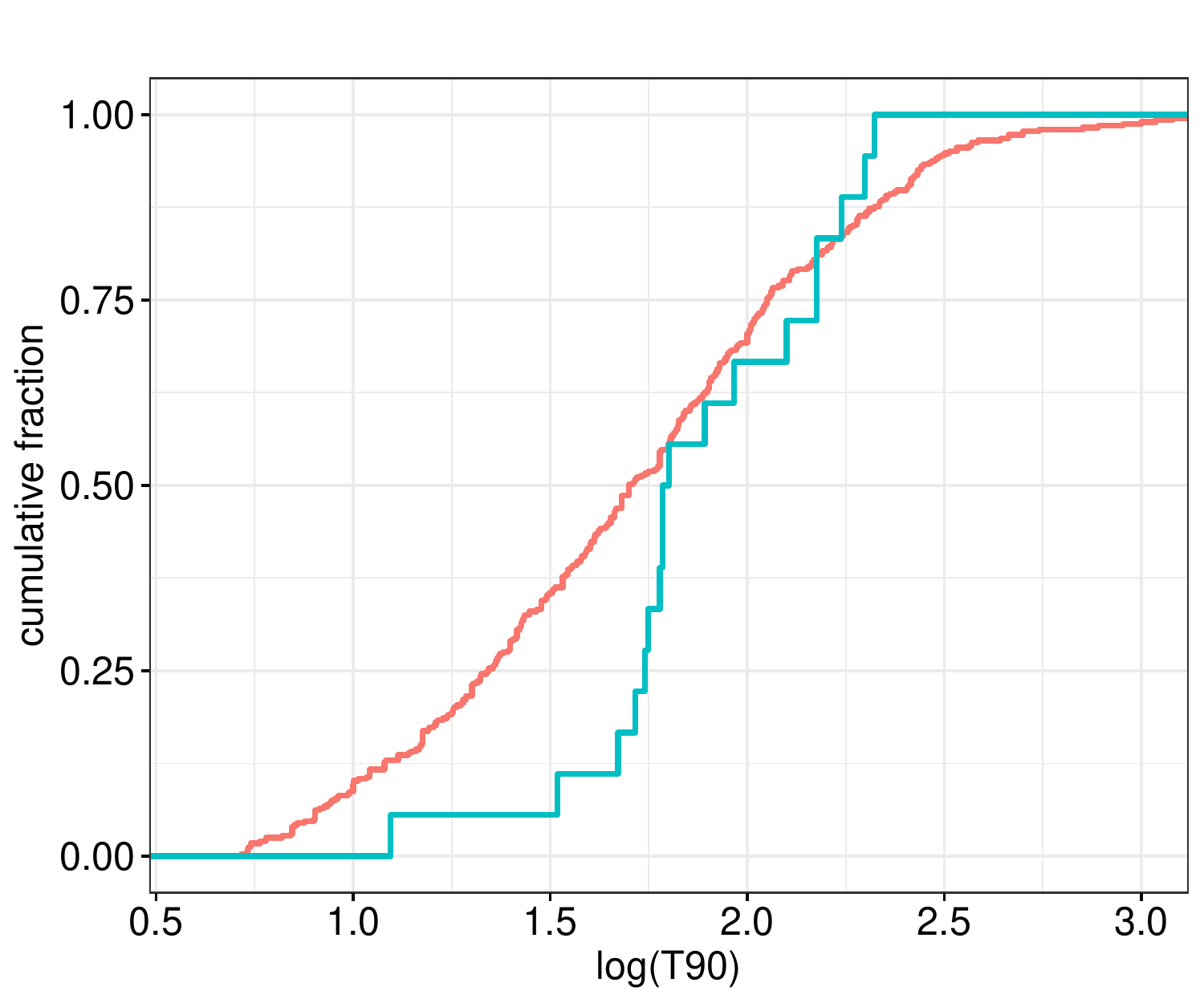}
   \caption{The GRBs between $1.49<z<1.61$ (AREA5, marked with blue line) are tend to be longer than the others (red line). }
  \label{fig:area5}
\end{figure}

\begin{figure}
 \centering
 \includegraphics[width=0.7\columnwidth]{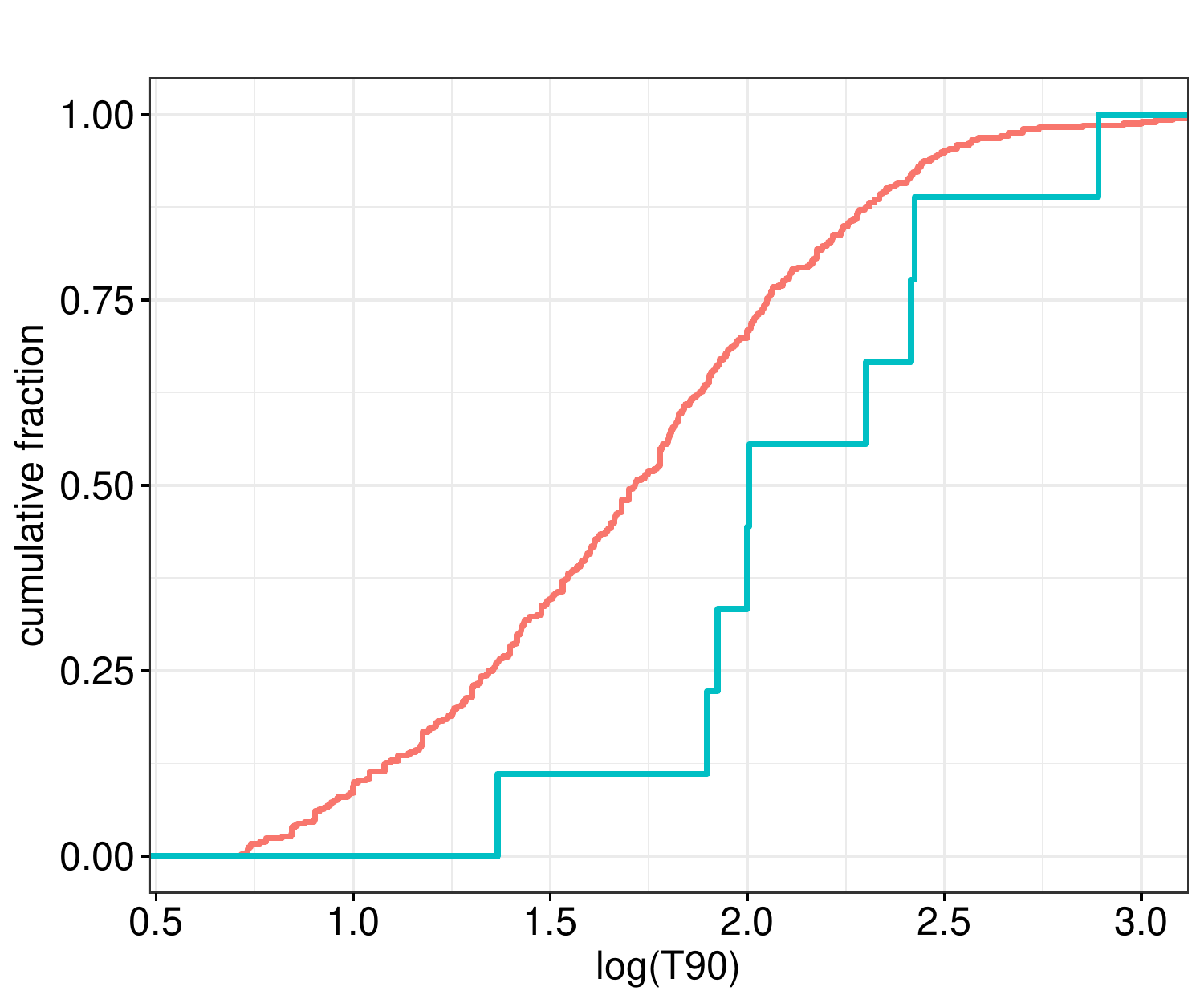}
   \caption{The GRBs between $2.91 < z < 3.075$ (AREA6, blue line)
    are also tend to be longer than the others (red line).}
  \label{fig:area6}
\end{figure}

Fig.~\ref{fig:figz0D4sik} shows the two parameter ($n$, redshift) $p$ value. The $p$ value reaches 0.0027 in two area. 
the $1.49<z<1.61$, $19<n<38$ (AREA5) and
$2.91<z<3.075$, $11<n<19$ (AREA6). 
Fig.~\ref{fig:area5}
and Fig.~\ref{fig:area6} show the cumulative redshift distribution of AREA5-6.
	
\begin{figure}
 \centering
 \includegraphics[width=0.7\columnwidth]{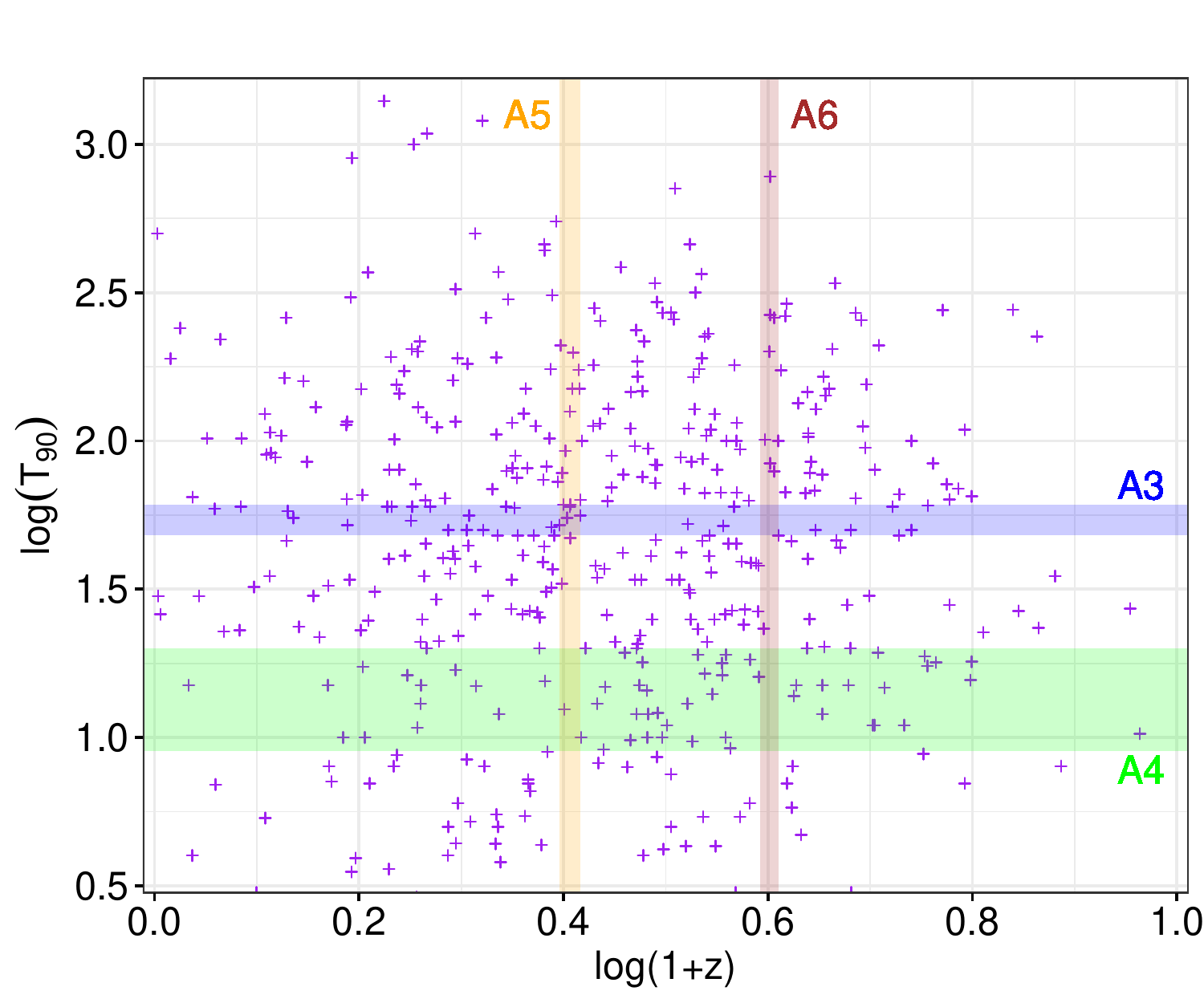}
  \caption{The four suspicious intervals: AREA3 is blue, AREA4 is green, AREA5 is yellow and AREA6 is brown.}
  \label{fig:figz0D21}
\end{figure}

\section{Summary \& conclusions}\label{sec:summary}

Several hundreds GRB redshift had been determined so far. Some
of the most important properties of the GRBs are the duration and the redshift of the bursts. In this paper we analysed 474 GRBs which had duration and redshift information as well. The most significant interrelation between these two quantities that the short GRBs redshifts are significantly smaller (we call this AREA1). Therefore, in this research we excluded the short GRBs (to be sure we choose the $T_{90}>5$s) and we analysed the remaining 421 GRBs.

We choose a certain number ($n=8-99$) GRBs which had a certain duration in an interval and compared their redshifts with the remaining GRBs redshift distribution. We find three interval where the redshift distribution was different than in the rest of the GRBs (see Fig.~\ref{fig:figarea3} and Fig.~\ref{fig:figarea4}). AREA2 ($T_{90}$ is between 16s, 20s and $n$ between 12 and 21) and AREA4 ($T_{90}$=(9s, 21s), $n$=(59, 68)) bursts are tend to be farther and AREA3 ($T_{90}$=(49s, 61s), $n$=(23, 36)) GRBs are tend to be closer than the average.
(It should be noted that the one-block KS significances will be lower as the whole sample because we do many, non-independent statistical tests.)

We also analysed the other way, chose a redshift interval and checked whether the $T_{90}$ distribution in this interval different than the remaining GRBs duration distribution (see Fig.~\ref{fig:area5} and Fig.~\ref{fig:area6}). We find two intervals where the duration distribution was tend to be longer than in the overall (see AREA5 and AREA6 in Fig.~\ref{fig:figz0D21}).

\authorcontributions{Conceptualization, Zsolt Bagoly, Lajos Balazs, Istvan Horvath, Sandor Pinter, and Istvan Racz; Data curation, Istvan Horvath, and Istvan Racz; Formal analysis, Sandor Pinter and Istvan Horvath; Investigation, Istvan Horvath, and Zsolt Bagoly; Methodology, Lajos Balazs, Istvan Horvath,  and Sandor Pinter; Project administration, Istvan Horvath, and Sandor Pinter; Resources, Zsolt Bagoly, Istvan Horvath, and Istvan Racz; Software, Sandor Pinter and Istvan Horvath; Supervision, Lajos Balazs, Zsolt Bagoly, Istvan Horvath, Sandor Pinter  and Istvan Racz; Validation, Zsolt Bagoly, Lajos Balazs, Istvan Horvath, Sandor Pinter and Istvan Racz; Visualization, Istvan Horvath and Sandor Pinter; Writing – original draft, Lajos Balazs, Zsolt Bagoly,  Sandor Pinter, Istvan Horvath,  and Istvan Racz; Writing – review \& editing, Zsolt Bagoly, Lajos Balazs, Sandor Pinter, Istvan Horvath,  and Istvan Racz.}

\funding{ Hungarian TKP2021-NVA-16 and OTKA K134257.}

\acknowledgments{
The authors thank the Hungarian TKP2021-NVA-16 and
OTKA K134257 program for their support.}
 
\conflictsofinterest{The authors declare no conflict of interest.} 
\end{paracol}
\reftitle{References}










\externalbibliography{yes}
\bibliography{horv22,references}

\end{document}